\documentclass[usenatbib]{mnras}
\usepackage{graphicx}
\usepackage{amstext}
\usepackage{amsmath}
\usepackage{natbib}
\usepackage{url}
\usepackage{setspace}
\usepackage{tabularx}
\usepackage{mathtools}
\usepackage[colorinlistoftodos]{todonotes}
\usepackage{booktabs}
\usepackage{rotating}

\def \araa {{\rm {ARA\&A}}}
\def \apj {{\rm {ApJ}}}

\def \apjs {{\rm {ApJS}}}
\def \apjl {{\rm {ApJL}}}

\def \aap {{\rm {A\&A}}}

\def \mnras {{\rm {MNRAS}}}

\def \pasp {{\rm {PASP}}}

\def     \Dntwoh {{\rm D$_\mathrm{frac}^\mathrm{N_2H^+}$}}
\def     \Dntwoheq {{\rm D$_\mathrm{frac, eq}^\mathrm{N_2H^+}$}}

\def	 \ntwoh {{\rm N$_2$H$^+$}}
\def     \ntwod {{\rm N$_2$D$^+$}}
\def	 \ceo	{{\rm C$^{18}$O}}

\def	 \ntwod {{\rm N$_2$D$^+$}}

\def 	 \kms {{\rm \,km\,s$^{-1}$}}
\def     \sol {{\rm M$_\odot$}}
\def     \arcsec {{\rm $^{\prime\prime}$}}
\def     \irdc {{\rm G035.39-00.33}}
\def     \micron{{\rm \,$\mu$m}}
\def     \ltsimm{\mathrel{\spose{\lower 3pt\hbox{$\sim$}}\raise 2.0pt\hbox{$<$}}}
\def     \gtsimm{\mathrel{\spose{\lower 3pt\hbox{$\sim$}}\raise 2.0pt\hbox{$>$}}}

\begin{document}

       

\title[Widespread deuteration throughout IRDC G035.39-00.33]{Widespread deuteration across the IRDC G035.39-00.33\thanks{Based on observations carried out with
    the IRAM 30m Telescope. IRAM is supported by INSU/CNRS (France),
    MPG (Germany) and IGN (Spain).}}  \author[Barnes, Henshaw, Caselli,  Kong, Jim\'{e}nez-Serra, Fontani, Tan]
     {A . T. Barnes$^{1,2}$\thanks{E-mail: astabarn@ljmu.ac.uk},
        S. Kong$^{3}$,
        J. C. Tan$^{3,4}$,
        J. D. Henshaw$^{1}$, 
        P. Caselli$^{5}$, \and 
        I. Jim\'{e}nez-Serra$^{6}$ and
        F. Fontani$^{7}$
        \\ {$^{1}$Astrophysics Research Institute, Liverpool John Moores University, 146 Brownlow Hill, Liverpool L3 5RF, UK }
       \\ {$^{2}$School of Physics and Astronomy, University of Leeds, LS2 9JT, Leeds, UK}
       \\ {$^{3}$Department of Astronomy, University of Florida, Gainesville, FL 32611, USA }
       \\ {$^{4}$Department of Physics, University of Florida, Gainesville, FL 32611, USA }
       \\ {$^{5}$Max-Plank-Institute for Extraterrestrial Physics (MPE), Giessenbachstrasse 1, 85748 Garching, Germany }
       \\ {$^{6}$University College London, Department of Physics and Astronomy, 132 Hampstead Road, London NW1 2PS, UK}
       \\ {$^{7}$INAF - Osservatorio Astrofisico di Arcetri, L.go E. Fermi 5, I-50125, Firenze, Italy }
       }

\date{Accepted 2016 February 18. Received 2016 February 15; in original form 2015 October 19.}
\pagerange{\pageref{firstpage}--\pageref{lastpage}} \pubyear{2015}

\maketitle

\label{firstpage}


\defcitealias{izaskun_2010}{Paper\,I}
\defcitealias{hernandez_2011}{Paper\,II}
\defcitealias{hernandez_2012a}{Paper\,III}
\defcitealias{henshaw_2013}{Paper\,IV}
\defcitealias{jimenez_2014}{Paper\,V}
\defcitealias{henshaw_2014}{Paper\,VI}

\begin{abstract}\label{abstract}

Infrared Dark Clouds (IRDCs) are cold, dense regions that are usually found within Giant Molecular Clouds (GMCs). Ongoing star formation within IRDCs is typically still deeply embedded within the surrounding molecular gas. Characterising the properties of relatively quiescent IRDCs may therefore help us to understand the earliest phases of the star formation process. Studies of local molecular clouds have revealed that deuterated species are enhanced in the earliest phases of star formation. In this paper we test this towards IRDC G035.39-00.33. We present an 80\arcsec \ by 140\arcsec \ map of the $J\,=\,2\rightarrow\,1$ transition of \ntwod, obtained with the IRAM-30m telescope. We find that \ntwod \ is widespread throughout G035.39-00.33. Complementary observations of \ntwoh\,($1-0$) \ are used to estimate the deuterium fraction, \Dntwoh\,$\equiv$\,$\mathrm{{\it N}(N_2D^+)/{\it N}(N_2H^+)}$. We report a mean \Dntwoh\,=\,0.04\,$\pm$\,0.01, with a maximum of \Dntwoh\,=\,0.09\,$\pm$\,0.02. The mean deuterium fraction is $\sim$\,3 orders of magnitude greater than the interstellar [D]/[H] ratio. High angular resolution observations are required to exclude beam dilution effects of compact deuterated cores. Using chemical modelling, we find that the average observed values of \Dntwoh are in agreement with an equilibrium deuterium fraction, given the general properties of the cloud. This implies that the IRDC is at least $\sim$\,3\,Myr old, which is $\sim$\,8 times longer than the mean free-fall time of the observed deuterated region. 
\end{abstract}



\section{Introduction}\label{introduction}

\begin{table*}
\caption{Frequency (MHz), Velocity Resolution,  Beam Size, and Beam $\&$ Forward Efficiency for the observed transitions. Note that the Forward and Beam Efficiencies have been extrapolated from the telescope specified values to the transition frequencies$^1$.}
\centering
{\large
\begin{tabular}{c c c c c c }
\hline
Transition  &$\nu$ (MHz) & $\Delta\nu_{res}$  & Beam Size$^1$ & Beam$^1$ & Forward$^1$ \\ [0.5ex]
 &  & (\kms)  & (arcsec) & Eff  & Eff \\ [0.5ex]

\hline

\multicolumn{1}{l}{\ntwod\,($2-1$)}  & 154217.18$^{a}$  & 0.31 & 16 & 0.66 & 0.93 \\
\multicolumn{1}{l}{\ntwoh\,($1-0$)$^{b}$}  & 93176.25$^{c}$ & 0.61 & 26 & 0.74 & 0.95 \\
\multicolumn{1}{l}{\ceo\,($1-0$)} & 109782.17$^{d}$ & 0.053 & 22 & 0.73 & 0.97 \\

\hline

\end{tabular}}

\begin{minipage}{\textwidth}
\smallskip
1: \url{www.iram.es/IRAMES/ mainWiki/Iram30mEfficiencies}. \\
$a$: Main hyperfine component, J=2-1, F$_1$= 3-2, F=4-3, from \citet{dore_2004}, with the central frequency updated to the value found in the CDMS catalogue (\url{www.astro.uni-koeln.de/cdms/catalog}). \\
$b$: Data from \citetalias{henshaw_2013}. \\
$c$: Isolated hyperfine component, J= 1-0, F$_1$= 0-1, F= 1-2, \citet{pagani_2009}. \\
$d$: \citet{cazzoli_2003}
\end{minipage}

\label{obs}
\end{table*}

Infrared Dark Clouds (IRDCs) are regions of cold (T $<$\, 25K; e.g. \citealp{ragan_2011}) and dense (n(H$_2$) $\geq$ 10$^{3-4}$\,cm$^{-3}$; e.g. \citealp{hernandez_2011, butler_2012}) gas, the most massive and dense of which have the potential to host the earliest stages of massive star and star cluster formation \citep{tan_2014}. Thus to understand these processes it is important to study the physical and chemical properties of IRDCs.

In such cold and dense environments, some molecules ``freeze-out'' onto the surfaces of dust grains forming icy mantles. Most notably CO is found to be highly depleted towards dense clouds within IRDCs (e.g. \citealp{fontani_2012a, giannetti_2014}), with line of sight CO depletion factors, $f_D$, of a few being reported towards the IRDC, \irdc \ (i.e. the observed gas phase abundance of CO is a few times smaller than expected in the case of no depletion; \citealp{hernandez_2011}).

Unlike CO, N-bearing species, in particular NH$_3$ and \ntwoh, better trace dense and cold gas (e.g. \citealp{caselli_1999, bergin_2002, bergin_tafalla_2007, difrancrsco_2007, fontani_2012b, fontani_2012c}). This is due to the fact that CO, largely frozen out, is unable to effectively destroy their molecular ion precursors (such as NH$^{+}$ and H$_{3}^{+}$), and hence less efficiently convert N$_2$H$^+$ into HCO$^{+}$. CO depletion, therefore, can boost the formation of nitrogen-bearing species. Furthermore, in the cold and dense environments of molecular clouds, abundances of deuterated nitrogen-bearing molecules are enhanced, as the formation rates of the deuterated forms of H$_{3}^{+}$ also increase with CO freeze-out (e.g. \citealp{dalgarno_1984, walmsley_2004}). These molecules are produced mainly by the exothermic proton-deuteron exchange reaction (for para-state reactants and products; \citealp{pagani_1992}): 

\begin{equation}
\mathrm{H_3^{+} + HD \rightarrow H_2D^{+} + H_2 + \Delta E},
\end{equation}
where $\Delta$E = 232\,K \citep{millar_1989}. For significant levels of CO depletion, temperatures below 30\,K, and for an ortho-to-para H$_{2}$ ratio less than $\sim\,0.1-0.01$ (e.g. \citealp{sipila_2013, kong_2015}), this reaction proceeds from left to right, producing an excess of H$_2$D$^+$ and an abundance ratio ([H$_2$D$^+$]/[H$^+_3$]) orders of magnitude larger than the interstellar [D]/[H] ratio ($\sim\,1.5\,\times\,10^{-5}$; e.g. \citealp{oliveira_2003}). Once the deuterated isotopologues of H$^+_3$ have formed, they can easily cede a deuteron to  other neutral species and enhance their abundances. For example, the reactions of deuterated isotopologues of H$^+_3$ with N$_2$ can produce \ntwod \ increasing the deuteration fraction of \ntwoh, \Dntwoh (we adopt the notation used by \citealp{kong_2015}, where the non-deuterated counterpart is shown in superscript).

Measurements of \Dntwoh towards low-mass cores range between $\simeq\,0.1-0.7$ (e.g. \citealp{crapsi_2004, belloche_2006, friesen_2010, friesen_2013, crapsi_2005, pagani_2007, bourke_2012}). For a sample of potential high-mass star forming regions, \citet{fontani_2006} found \Dntwoh\,$\simeq$\,0.015. \citet{miettinen_2011} found \Dntwoh\,$\simeq$\,0.002-0.028 toward massive clumps within several IRDCs, while deuterium fractions as high as in low-mass prestellar cores have been observed toward massive starless clumps/cores embedded in quiescent IRDCs \citep{fontani_2011}.

This paper focusses on the massive (17,000\,$\pm$\,5000\,\sol; \citealp{kainulainen_2013}), filamentary IRDC, \irdc \ which resides at a distance of 2.9\,$\pm$\,0.5\,kpc \citep{simon_2006b}\footnote{Following \citet{hernandez_2011}, we adopt the kinematic distances of \citet{simon_2006b}, who assumed the \citet{clemens_1985} rotation curve. This leads to a distance of 2.9\,kpc for \irdc. The uncertainties in this distance are likely to be of order 0.5\,kpc, which could result, for example, from line-of-sight noncircular motions of $\sim$\,8\kms.}. This IRDC was first identified as having the potential to host massive cluster formation by \citet{rathborne_2006}. The extinction mapping of \citet{butler_2012} found several high mass surface density ``cores'' within \irdc, most notably the ``H6'' region which contains $\sim$\,70\sol \ of material within a radius of $\sim$\,0.25\,pc. In recent years this IRDC has been the subject of an in-depth analysis, which has focussed on: large-scale shocks traced by SiO emission (\citealp{izaskun_2010}, \citetalias{izaskun_2010},), widespread CO depletion (\citealp{hernandez_2011}, \citetalias{hernandez_2011}), the virial state of the cloud (\citealp{hernandez_2012a}, \citetalias{hernandez_2012a}) and the kinematics of the low and high density gas and the excitation conditions in the cloud at varying scales (\citealp{henshaw_2013, jimenez_2014, henshaw_2014}, \citetalias{henshaw_2013, jimenez_2014, henshaw_2014}, respectively). As widespread CO depletion has been found in \irdc, this cloud is an ideal candidate to also exhibit widespread deuteration. This paper presents IRAM-30m \ntwod\,($2-1$) observations of \irdc, with the aim of estimating the column density and deuterium fraction, and ultimately finding the evolutionary state of the cloud.


\section{Observations}\label{observations}

The \ntwod\,($2-1$) observations were carried out throughout August 2009 with the Institut de Radioastronomie Millim\'etrique 30-m telescope (IRAM-30m) at Pico Veleta, Spain. The large-scale images were obtained in the On-The-Fly (OTF) mapping mode. The central coordinates of the maps are $\alpha$(J2000)\,=\,18$^h$57$^m$08$^s$, $\delta$(J2000)\,=\,02$^{\circ}$10$'$30${''}$ (\textit{l}\,=\,35.517$^{\circ}$, \textit{b}\,=\,-0.274$^{\circ}$). The off-source position used was (300\arcsec, 0\arcsec; in relative coordinates). The EMIR receivers were used. The VErsatile SPectrometer Assembly (VESPA) provided a spectral resolution at 156\,kHz (equivalent to 0.3 \kms) at the frequency of the \ntwod($2-1$) line (main hyperfine component frequency 154217.1805\,MHz; \citealp{dore_2004}). The data were converted into main beam brightness temperature, T$_{\mathrm{MB}}$, from antenna temperature, T$_{\mathrm {A}}^{*}$, by using the beam and forward efficiencies shown in Table\,\ref{obs}. Saturn was observed to calculate the focus, and pointing was checked every $\sim$\,2\,hours on G34.3+0.2. The data were calibrated with the chopper-wheel technique \citep{kutner_1981}, with a calibration uncertainty of $\sim$\,20\%. Information on the beam sizes, frequencies, velocity resolutions are summarised in Table\,\ref{obs}. 

{\sc gildas}\footnote{see \url{https://www.iram.fr/IRAMFR/GILDAS/}} packages {\sc class} and {\sc mapping} were used to post-process the data. This included subtracting a single- order polynomial function to produce a flat baseline, and convolving the OTF-data with a Gaussian kernel, increasing the signal- to-noise ratio, and allowing us to resample the data onto a regularly spaced grid. The absolute angular resolution of the IRAM-30m antenna at the frequency of the J\,=\,$2\,\rightarrow\,1$ transition of \ntwod \ is $\sim$\,16\arcsec. Throughout this work, all line data are spatially smoothed to achieve an effective angular resolution of $\sim$\,27\arcsec, with a pixel spacing of 13.5\arcsec, to allow comparison with the \ntwoh data.

We utilise the \ntwoh\,($1-0$) map from \citetalias{henshaw_2013} (see this paper for more information on the \ntwoh observations), CO depletion map of \citetalias{hernandez_2012a}, and the mass surface density map from \citet{kainulainen_2013}.


\begin{figure}
\centering
\includegraphics[trim = 0mm 0mm 0mm 0mm, clip,width=0.84\columnwidth]{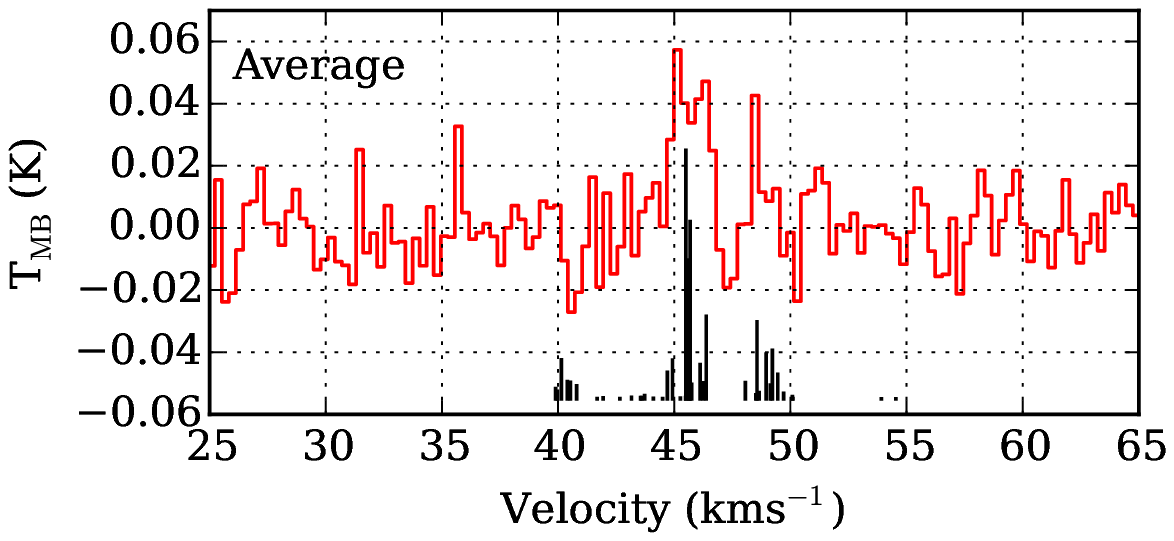}
\includegraphics[trim = 0mm 0mm 0mm 0mm, clip,width=0.84\columnwidth]{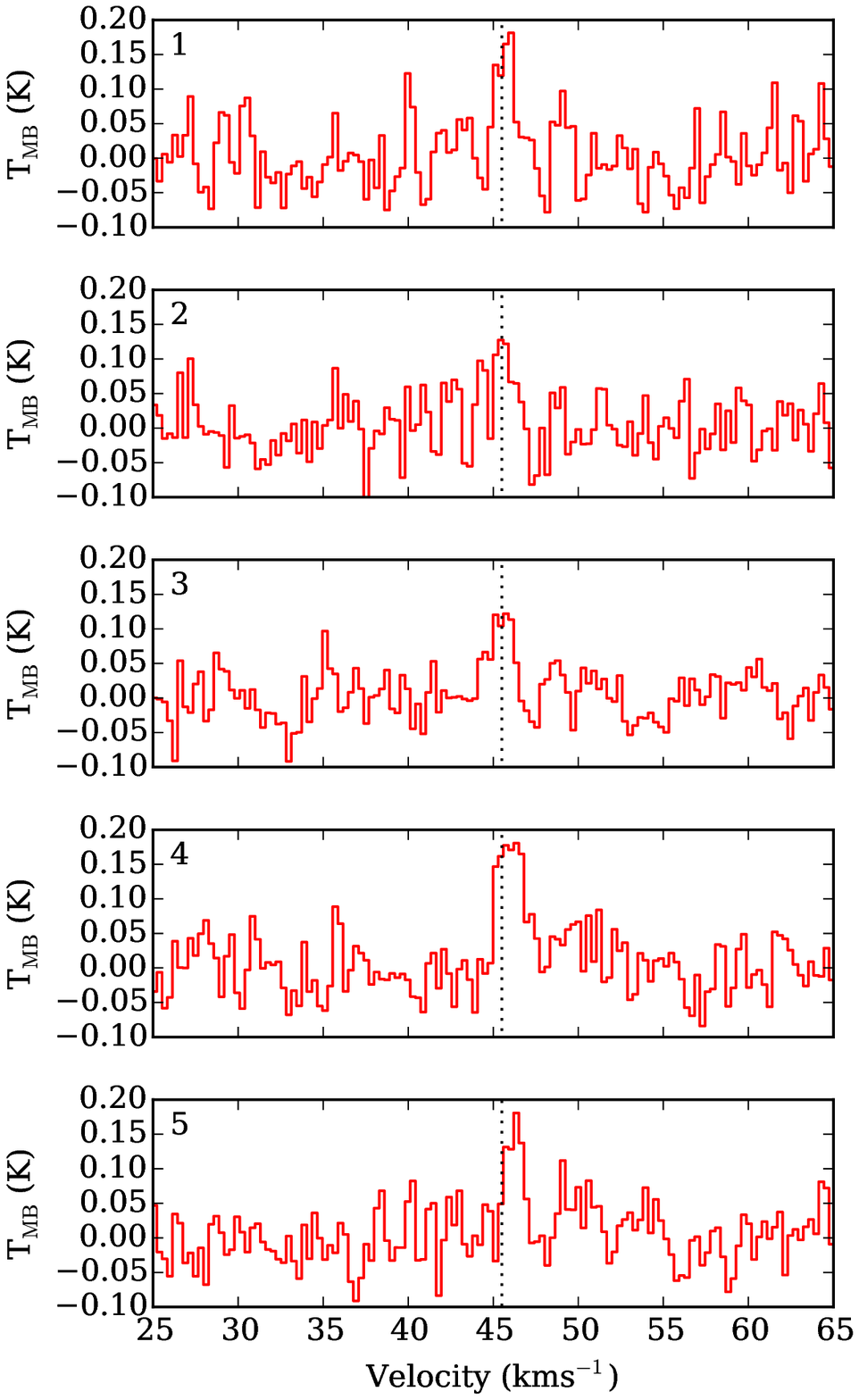}
\caption{The upper panel shows the average spectrum of \ntwod\,($2-1$) for all positions (red histogram). The vertical black lines below the spectrum indicate the positions and relative strengths of all the hyperfine components of the \ntwod\,($2-1$) transition \citep{dore_2004}, assuming a rest velocity of 45.5\kms. The lower panels are example spectra taken from positions of high integrated intensity, represented as the boxes on Figure\,\ref{n2dp21-II}. The dotted lines represent a velocity of 45.5\,\kms.}
\label{spec}
\end{figure}

\section{Results}\label{results}

\begin{figure*}
\begin{center}
\includegraphics[trim = 0mm 0mm 0mm 0mm, clip,width=1.7\columnwidth]{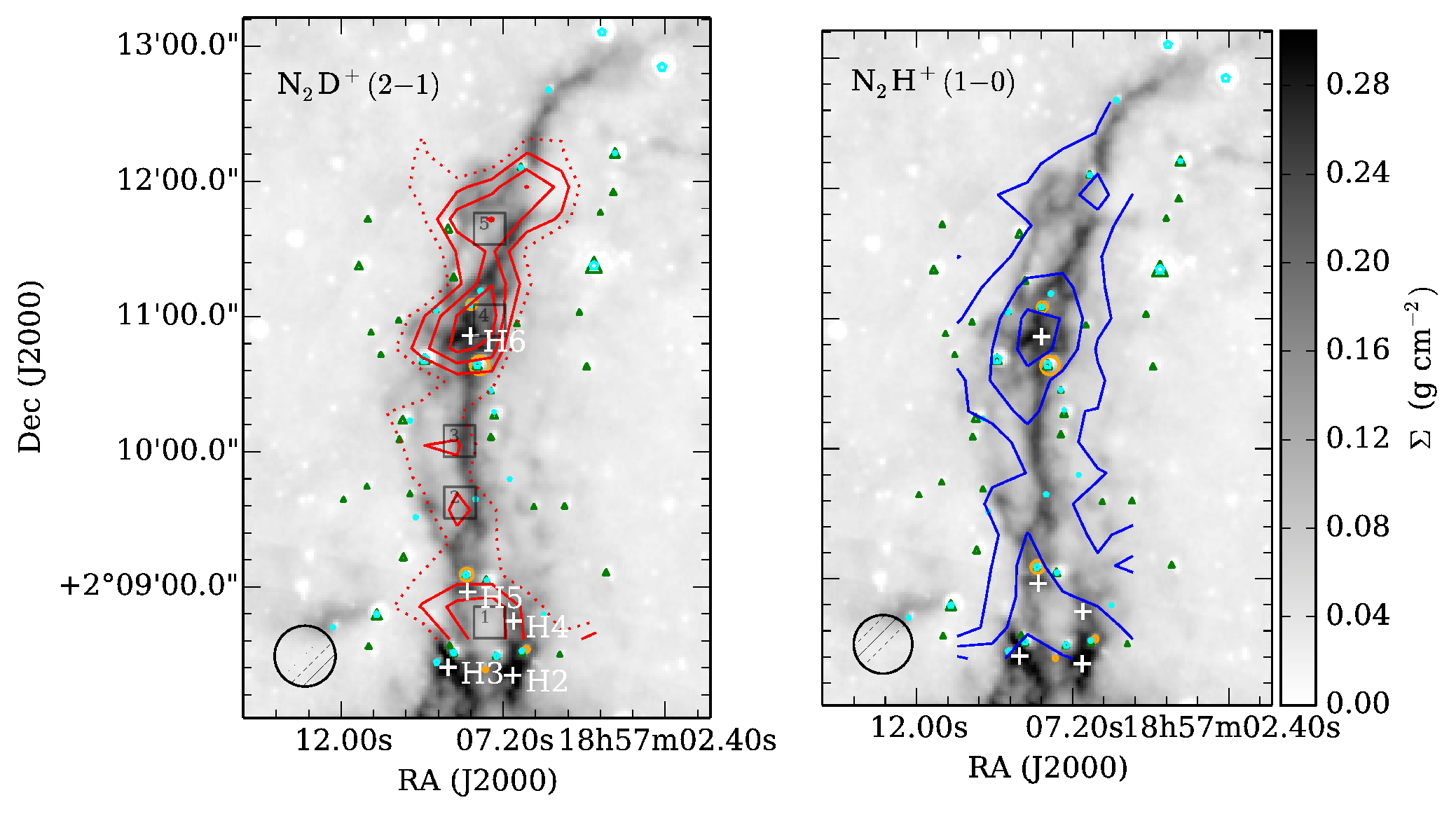}
\end{center}
\caption{The left panel shows the integrated intensity map of \ntwod\,($2-1$) emission seen in contours of 2$\rm{\sigma}$ (dotted contour), 3$\rm{\sigma}$, 4$\rm{\sigma}$ and 5$\rm{\sigma}$; where $\rm{\sigma}$\,=\,0.07\,K\,\kms. Overlaid are boxes which represent the positions of the spectra shown in Figure\,\ref{spec}. The right panel shows the integrated intensity contours of the \ntwoh\,($1-0$) emission, data taken from \citetalias{henshaw_2013}; contours are 5, 10 and 15\,$\sigma$, where $\sigma\,\sim\,0.11$\,K\,\kms. Each contour map is overlaid on the mass surface density map \citep{kainulainen_2013}. White crosses indicate the positions of the massive cores \citep{butler_2012}, and the symbols indicate the positions of the 8\micron \ and 24\micron \ sources (green triangles and cyan pentagons, respectively; \citealp{carey_2009}), and the 4.5\micron \ extended emission sources or `green fuzzies' (orange circles; \citealp{chambers_2009}). The size of each source represents the strength of the emission \citepalias{izaskun_2010}. The hatched circle represents the effective spatial resolution of the observations (after smoothing).}
\label{n2dp21-II}
\end{figure*}

Figure\,\ref{spec} presents the averaged spectrum for the mapped region, and some example spectra taken at positions of high integrated intensity (see Figure\,\ref{n2dp21-II}). All emission above a 3\,$\sigma$ level of $\sim$\,0.21\,K\,\kms ($\sigma$\,=\,$\sigma_\mathrm{RMS}$\,$\Delta \nu_\mathrm{res} \,\sqrt{\mathrm{N}_\mathrm{ch}}$; where $\sigma_\mathrm{RMS}$ is the root mean square noise of the spectrum in K, $\mathrm{N}_\mathrm{ch}$ is the number of channels and $\Delta \nu_\mathrm{res}$ is the velocity resolution in \kms), is seen within a velocity range of $40-50$\,\kms. This range is comparable to that found by the previously mentioned works on this cloud, and therefore we can be confident that the observed \ntwod \ emission is associated with \irdc. 

Each spectrum has been inspected by-eye for the presence of multiple velocity components previously identified in \ntwoh \ and \ceo \ emission (\citetalias{henshaw_2013}; with mean \ntwoh \ emission velocities of 42.95\,$\pm$\,0.17\kms, 45.63\,$\pm$\,0.03\kms, and 46.77\,$\pm$\,0.06\kms). However, evidence of only one component, centred at $\sim$\,46\,\kms \ was found in the \ntwod \ data. Unlike the \ntwoh\,($1-0$) line, where one of the seven hyperfine components is isolated, \ntwod\,($2-1$) possesses no isolated hyperfine components; the line is a blend of 40 hyperfine components, spread across a velocity range of 14.6\,\kms \ (as shown in Figure\,\ref{spec}, upper panel). This, as well as linewidths of $\sim$\,1\,\kms \ (i.e. similar to those of \ntwoh\,($1-0$); \citetalias{henshaw_2013}), makes the identification of multiple velocity components very difficult.

Figure\,\ref{n2dp21-II} presents a map of the \ntwod\,($2-1$) emission integrated between $40-50$ \kms. The integrated intensity contours are overlaid on the mass surface density map of \citet{kainulainen_2013}, and superimposed are the positions of the massive cores first identified by \citet{rathborne_2006} in millimetre continuum emission, which have been repositioned to the mid-infrared extinction peaks by \citet{butler_2012}. In Figure\,\ref{n2dp21-II} we also show the location and strength of the 4.5\micron \ emission (`green fuzzies'; \citealp{chambers_2009}), and the 8\micron \ and 24\micron \ \citep{carey_2009}. The \ntwod \ emission is concentrated towards the ``H6'' region, and the south, near H2, H3, H4 and H5 ``core'' regions. However, when considering the 2\,$\sigma$ emission level there is evidence to suggest that the emission is extended across a large portion of the cloud.


\section{Analysis}\label{analysis}

\subsection{Column density} \label{column density}

The column densities are calculated from the integrated intensity of the \ntwod\,($2-1$) line, following the procedure outlined in \citet{caselli_2002} for optically thin emission. This is a reasonable assumption based on the relatively faint lines detected across the cloud. We assume a constant excitation temperature of $T_\mathrm{ex}\,\sim4.5$\,K (equivalent to the mean $T_\mathrm{ex}$ derived from the \ntwoh \ observations; \citetalias{henshaw_2013}). Note, that ranging the $T_\mathrm{ex}$ between 4-20\,K would cause the column density to vary by $N$\,(\ntwod)$^{+30\%}_{-60\%}$. 

The mean beam-averaged column density is $N$\,(\ntwod)\,=\,6.2\,$\pm$\,1.4\,$\times$\,10$^{11}$\,cm$^{-2}$, when imposing a three sigma threshold on the \ntwod\,($2-1$) emission. The maximum beam-averaged column density of \ntwod \ is found toward the H6 region, with a value of $N$\,(\ntwod)\,=\,8.0\,$\pm$\,1.4\,$\times$\,10$^{11}$\,cm$^{-2}$.

\subsection{Deuterium fraction} 

\begin{figure}
\begin{center}
\includegraphics[trim = 0mm 0mm 0mm 0mm, clip,width=0.85\columnwidth]{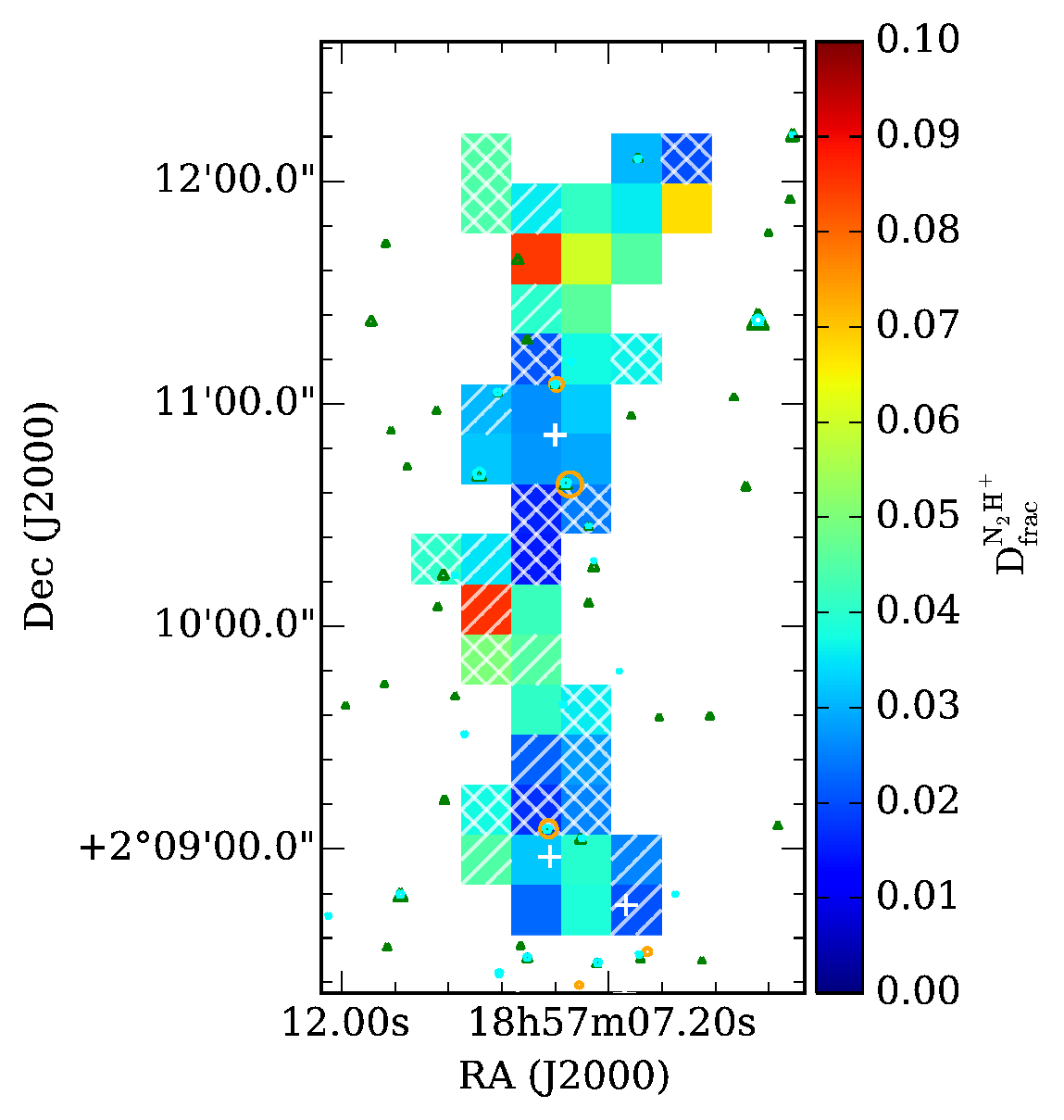}
\end{center}
\caption{A map of the deuterium fraction. This map includes the emission of both \ntwod \ and \ntwoh \ detected at a 2\,$\sigma$ level shown cross hatched, above a 2.5\,$\sigma$ level shown hatched, and above 3\,$\sigma$ with no hatch. Symbols are identical to those in Figure\,\ref{n2dp21-II}.}
\label{d_frac}
\end{figure}

The \ntwod \ to \ntwoh \ column density ratio is used to define the deuterium fraction across the mapped region. Figure\,\ref{d_frac} shows the deuterium fraction for positions where the emission of both \ntwod \ and \ntwoh \ is detected above a 2\,$\sigma$ (cross hatched), above a 2.5\,$\sigma$ (hatched), and above a 3\,$\sigma$ (no hatch) level. We find values of the deuteration fractions larger than 0.01 widespread throughout the cloud, with the highest values found towards the north. 

Taking into account only the emission above 3\,$\sigma$, the mean beam-averaged deuterium fraction across \irdc \ is \Dntwoh\,=\,0.04\,$\pm$\,0.01. The maximum value is found north of the ``H6'' region (at $\alpha$(J2000)\,=18$^h$57$^m$09$^s$,  $\delta$(J2000)\,=\,02$^{\circ}$11$'$39${''}$), with a value of \Dntwoh\,=\,0.09\,$\pm$\,0.02. Note that varying the excitation temperature for both \ntwoh \ and \ntwod \ between 4-20\,K would cause would cause a change of \Dntwoh$^{+15\%}_{-55\%}$.

\begin{figure}
\begin{center}
\includegraphics[trim = 0mm 0mm 0mm 0mm, clip,width=1\columnwidth]{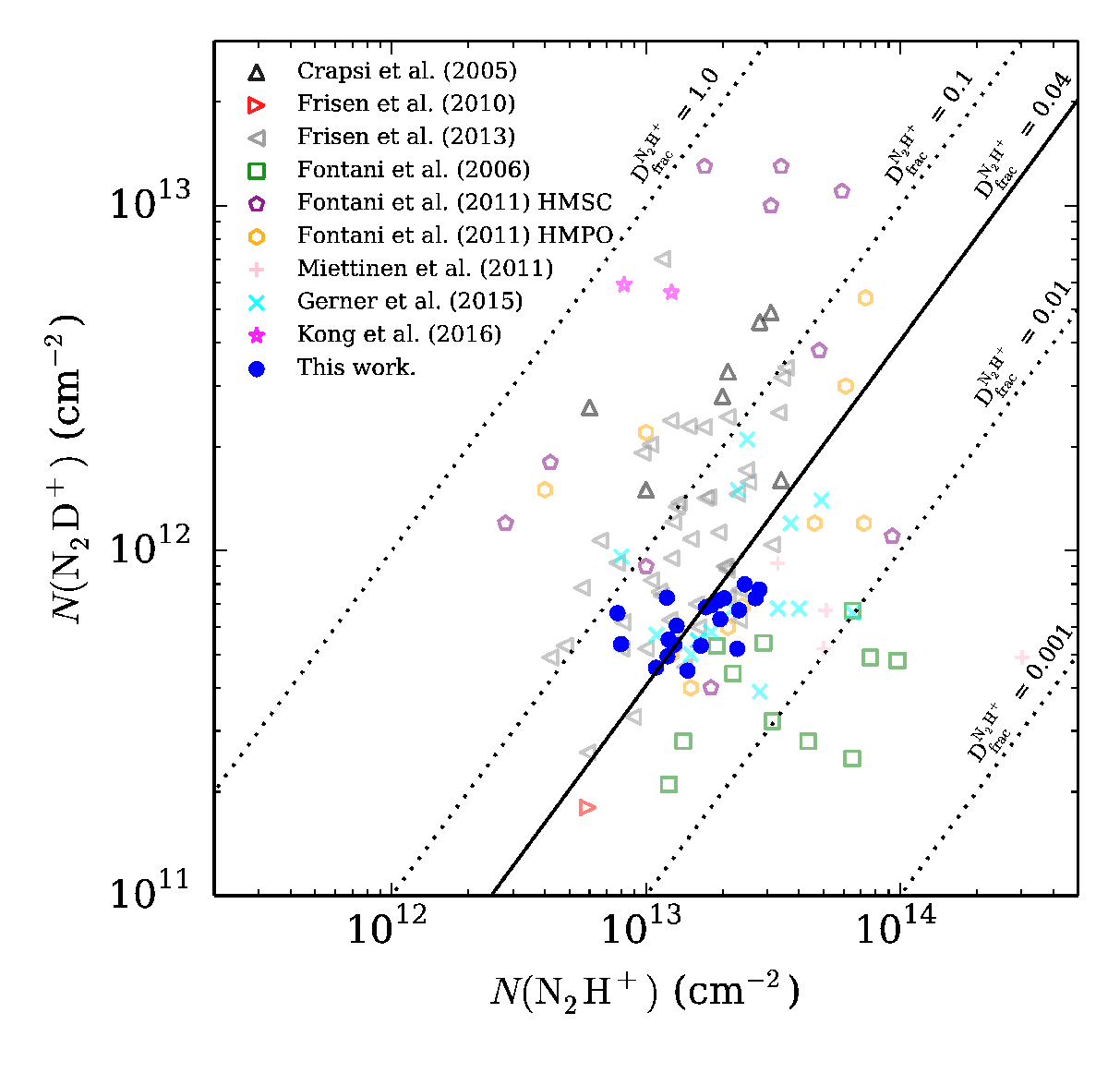}
\end{center}
\caption{Column density of \ntwoh \ as a function of the column density of \ntwod, for the values found in this work and those found within the literature; triangles represent the low-mass star forming regions, the additional shapes represent high-mass star forming regions. Note that only the high-mass protostellar objects and starless clumps (HMPO and HMSC, respectively), and IRDCs are plotted from the surveys of \citet{fontani_2011} and \citet{gerner_2015}. The lines overlaid represent deuterium fractions of \Dntwoh\,=\,[0.001, 0.01, 0.04, 0.1, 1.0] from right to left, respectively.}
\label{n2h_vs_n2d}
\end{figure}

\section{DISCUSSION}\label{discussion}

\subsection{G035.39-00.33 in the context of other star-forming regions}\label{comparison_works}

As in low-mass star-forming regions, the deuterium fraction of massive star-forming regions is believed to be a good evolutionary tracer (e.g. \citealp{fontani_2011}). However, the exact boundaries to define an evolutionary state are still poorly understood. To show this, plotted on Figure\,\ref{n2h_vs_n2d} is the column densities of \ntwod \ and \ntwoh \ found in this work, and for various other high-mass (\citealp{fontani_2006, fontani_2011, miettinen_2011, gerner_2015, kong_2016}) and low-mass (triangles only; \citealp{crapsi_2005, friesen_2010, friesen_2013}) cores described within the literature. 

We find that our deuterium fractions are most similar to those found in IRDCs by \citet{gerner_2015}, who found mean values of $\sim$\,0.01-0.1 (observed at spatial resolution of a $\sim$\,few 10\arcsec, at $\sim$\,4\,kpc). However, our values are an order of magnitude lower than the values observed by \citet{fontani_2011} and \citet{kong_2016} towards high-mass starless cores (HMSCs), where \Dntwoh\,$\sim$\,0.3. \citet{gerner_2015} suggest that the low deuteration values found in their sample of IRDCs may be due to the presence of unresolved evolved objects (24\micron \ sources). This could also be the case for \irdc, as the average deuterium fraction is also close to the values observed towards high-mass protostellar candidates ($\sim$\,0.04; \citealp{fontani_2006, fontani_2011}).

\begin{figure*}
\begin{center}
\includegraphics[trim = 0mm 0mm 0mm 0mm, clip,width=1.95\columnwidth]{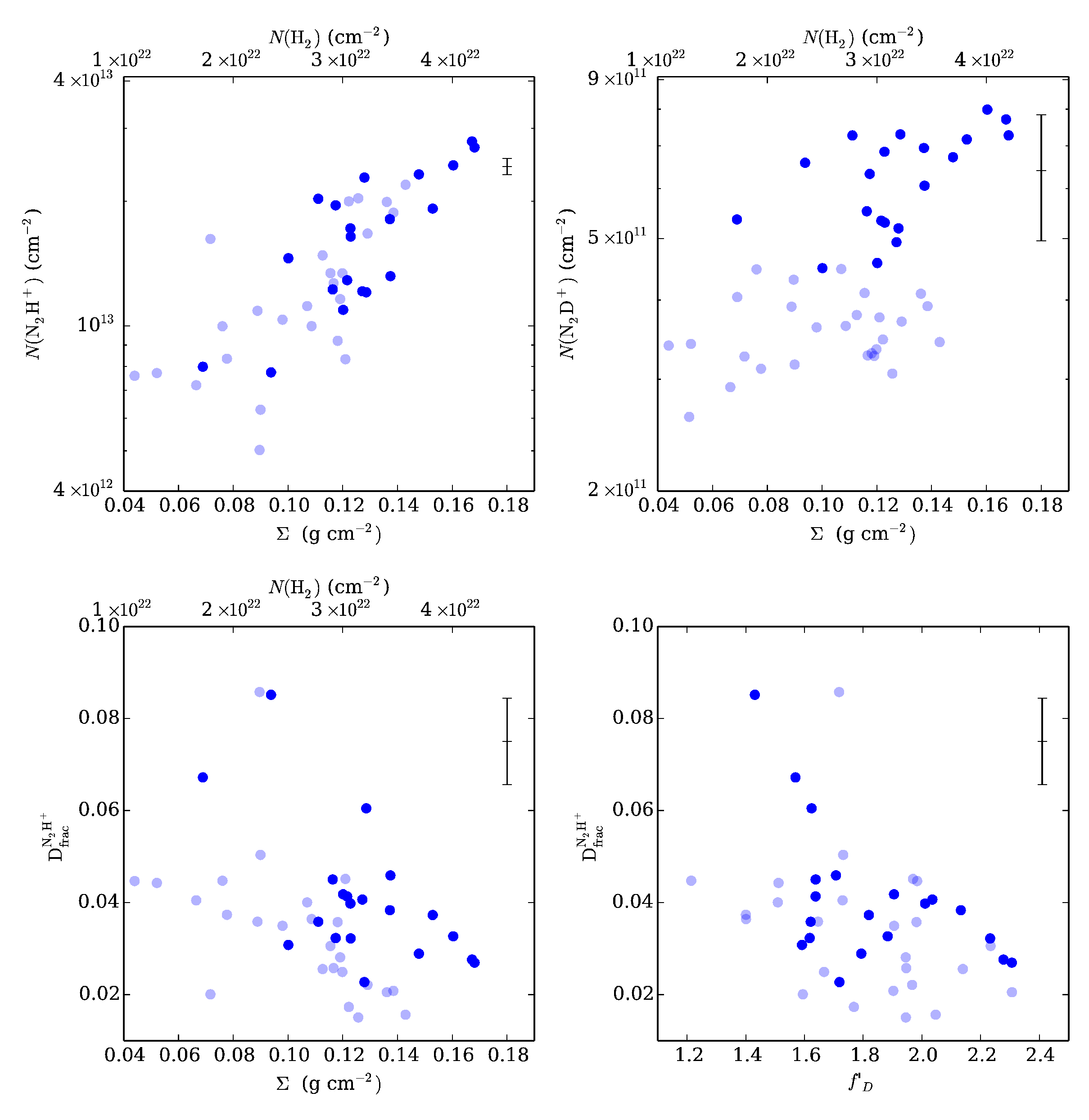}
\end{center}
\caption{Shown in the panels are the \ntwoh \ (upper left) and \ntwod \ (upper right) column densities as a function of mass surface density, deuterium fraction as a function of mass surface density (lower left), and deuterium fraction as a function of normalised CO depletion factor (lower right). Average errors for \ntwod column density and \Dntwoh \ are displayed in the upper right of each plot. Not shown are the errors on the mass surface density and CO depletion, which are $\sim$30\% \citep{kainulainen_2013} and $\sim$50\% (\citetalias{hernandez_2012a}), respectively. The solid and transparent points represent positions where both the \ntwoh \ and \ntwod \ emission is above a 3$\sigma$ and 2$\sigma$ error threshold, respectively.}
\label{corr}
\end{figure*}

To investigate how the \ntwoh and \ntwod \ emission and the deuterium fraction vary within \irdc, we plot them as a function of mass surface density in Figure\,\ref{corr}. If we consider only the positions above a 3$\sigma$ error threshold, the column density of \ntwod \ remains relatively constant with increasing mass surface density (dynamical range of $\sim$\,1.5, which is similar to the scale of the uncertainties). However, for the same positions, the column density of \ntwoh \ shows a significant positive gradient with increasing mass surface density (dynamical range of $\sim$\,3). This is reflected in the plot of deuterium fraction as a function of mass surface density, which shows an overall negative correlation. This is consistent with a picture in which the \ntwod \ is more spatially concentrated in cores than the \ntwoh, which is also present in the clump envelope (that dominates the mass).

Observations show that the deuterium fraction is highly sensitive to the level of CO freeze-out (e.g. \citealp{caselli_2002a}), where the CO depletion factor, $f_{D}$, can be expressed as the ratio of the observed mass surface density to mass surface density derived from CO emission, assuming a reference CO fractional abundance with respect to H$_2$. \citetalias{hernandez_2012a} calculated the CO depletion averaged along each line-of-sight (i.e. each pixel) throughout \irdc, $f'_{D}$, which is normalised such that on average pixels with mass surface densities of 0.01\,g\,cm$^{-2}$$<\Sigma<$0.03\,g\,cm$^{-2}$ are unity. Figure\,\ref{corr} displays the deuterium fraction as a function of normalised CO depletion factor, for which we do not see a positive correlation, but rather again an anti-correlation. This is contrary to what is expected, that level of deuteration is elevated in dense, CO depleted cores; but it is reminiscent of the results recently obtained toward the Ophiuchus low-mass star forming region \citep{punanova_2015}. These authors suggest that the highly deuterated but CO-rich cores may be recently formed, centrally concentrated starless cores.

These results could indicate that along the line-of-sight, high mass surface density and CO depleted positions have enhanced \ntwoh, but the \ntwod \ is only tracing a portion of this in the 3rd dimension. We, therefore, suggest that measurements of the deuterium fraction in massive star-forming regions could be limited by low spatial resolution observations (beam dilution), and/or the relatively unknown evolutionary stage of the gas (whether or not it has reached chemical equilibrium or is being influenced by the presence of YSOs). These are discussed in more detail in the following section.

\subsection{Comparison with chemical models}\label{comparison}

To determine if the observed levels of deuteration are consistent with the current evolutionary stage of the \irdc, we have conducted a series of chemical models \citep{kong_2015}. The model consists of the Nahoon code and a reduced network extracted from KIDA database \citep{wakelam_2012}, including the elements H, D, He, C, N, O. The chemical species traced by the code contain up to 3 atoms in size, except for $\rm H_3O^+$ and its deuterated isotopologues, which significantly improve the consistency with a more complete network \citep{sipila_2013}. Spin states of $\rm H_2$, $\rm H_3^+$ and their deuterated isotopologues are included, and the formation of $\textrm{o-H}_2$, $\textrm{p-H}_2$, HD, $\textrm{o-D}_2$, $\textrm{p-D}_2$ on dust grain surface are considered following \citet{lepetit_2002}. We follow \citet{pagani_2009b} in calculating the dissociative recombination rates of all forms of $\rm H_3^+$. The initial elemental abundances are listed in table 2 in \citet{kong_2015}. In this paper, we treat the depletion of neutral species by reducing the initial elemental abundances of C, N, O by the depletion factor, $f_{\rm D}$, and consider a broad combination of physical conditions appropriate for IRDC G035.39-00.33. 

\begin{figure}
\begin{center}
\includegraphics[trim = 0mm 0mm 0mm 0mm, clip,width=1\columnwidth]{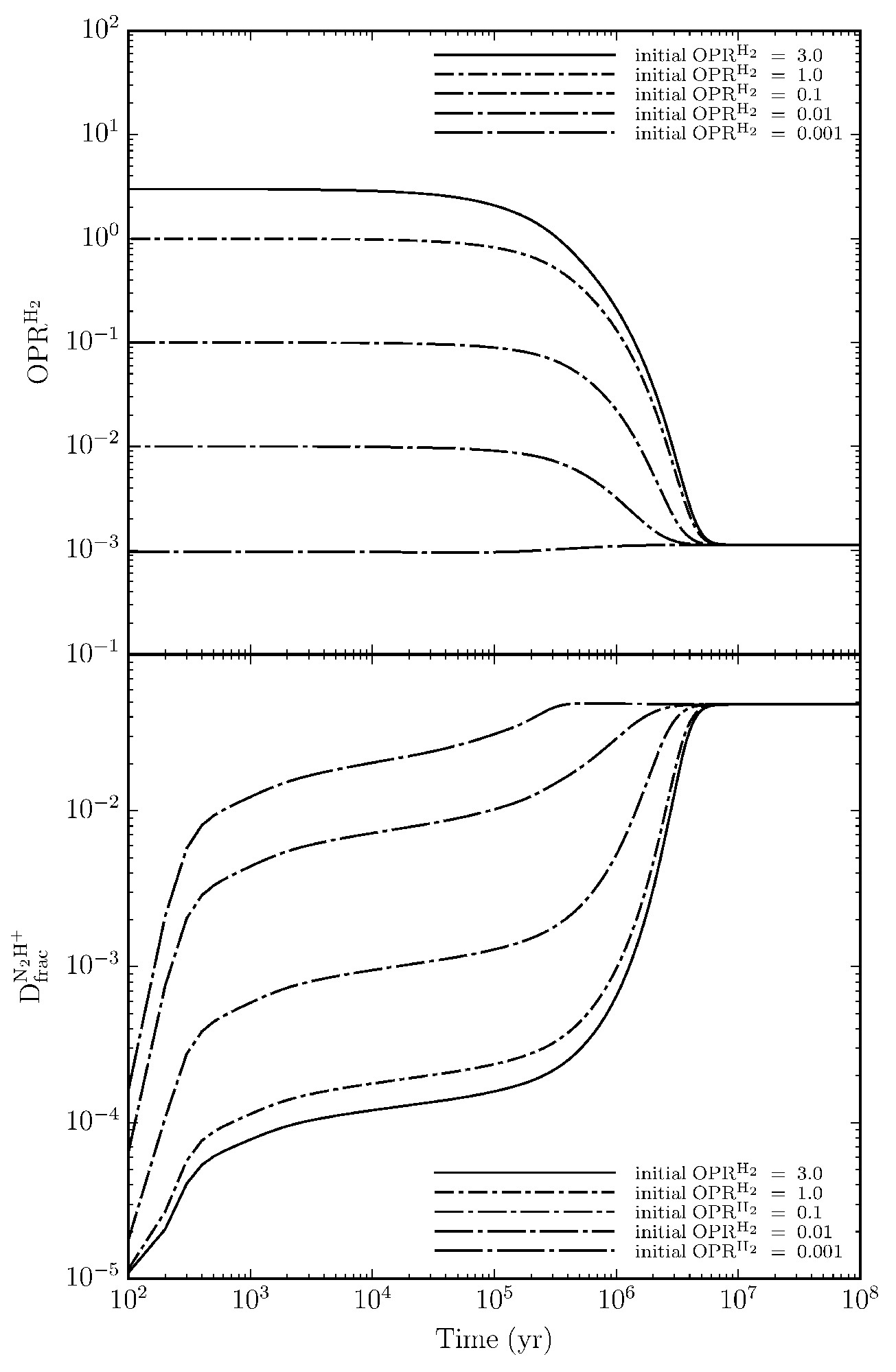}
\end{center}
\caption{Time evolution of the OPR$^{\rm H_2}$ (upper panel) and \Dntwoh (lower panel) under different assumptions of the initial OPR$^{\rm H_2}$, for $n_\mathrm{H}$\,=\,10$^{4}$\,cm$^{-3}$, $f_D$\,=\,3, T$_{kin}$\,= \,15\,K and A$_{v}$\,=\,20\,mag. We explore OPR$^{\rm H_2}$ from 3 down to 0.001.}
\label{models}
\end{figure}

%

\begin{figure}
\begin{center}
\includegraphics[trim = 0mm 0mm 0mm 0mm, clip,width=1\columnwidth]{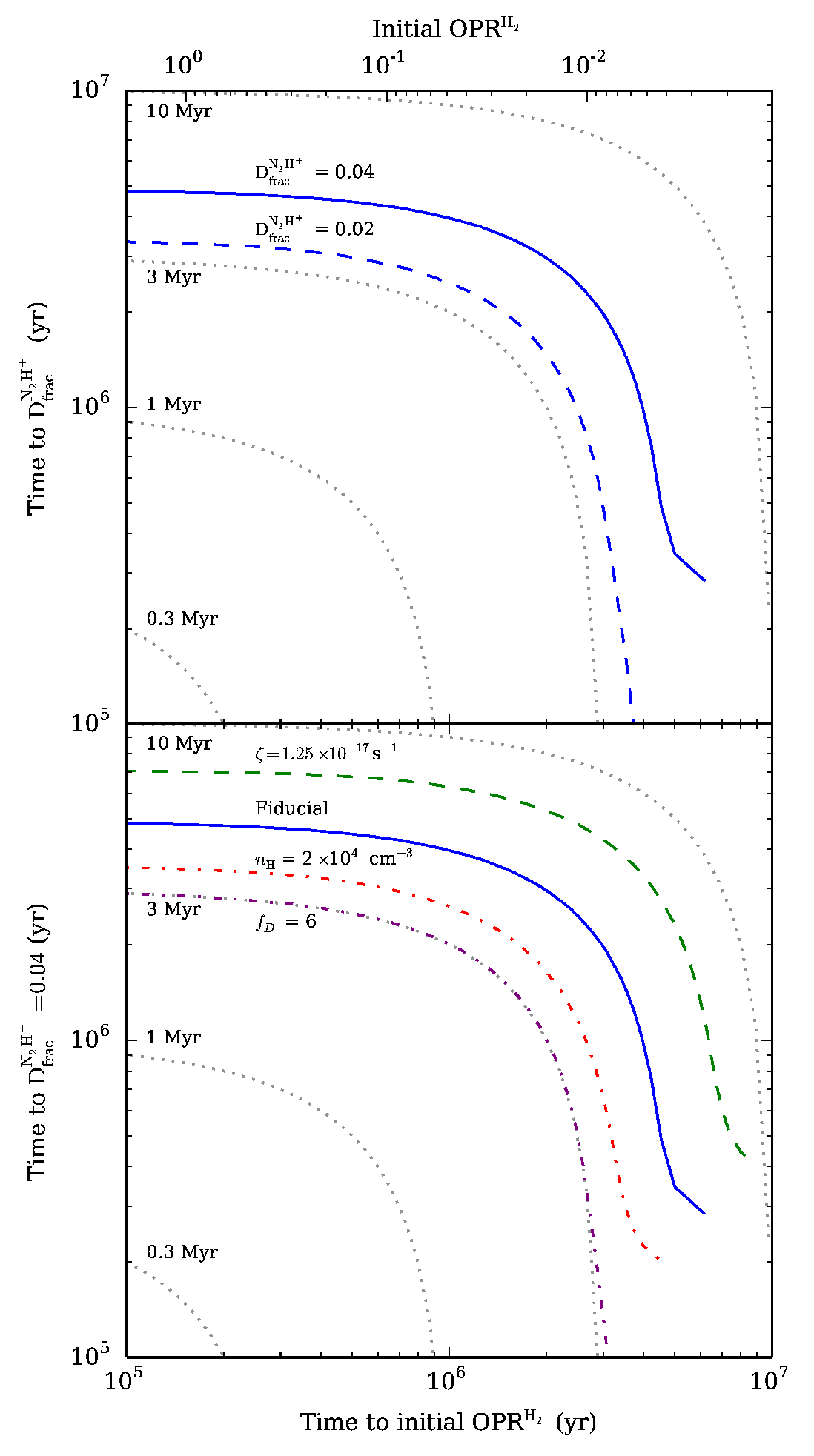}
\end{center}
\caption{Time to reach the observed deuterium fraction (\Dntwoh\,=\,0.04) from a given initial ortho-to-para ratio of H$_2$ (shown on the top axis) versus the time to reach this initial ortho-to-para ratio if starting from statistical equilibrium ratio of 3. The solid blue line shows this result for the fiducial model parameters, e.g. of density, temperature, depletion factor, cosmic ray ionisation rate. Grey dotted lines show contours of the sum of the deuteration timescale and ortho-to-para ratio decay timescale to equal 0.3, 1, 3, 10\,Myr, as labelled. The dashed blue line (upper panel) shows the time to reach \Dntwoh\,=\,0.02 for the fiducial model. The dashed and dot-dashed lines (lower panel) show the effect of varying several of the model parameters, as labelled.}
\label{models-OPR}
\end{figure}

We explore a grid of models with A$_{v}$ = [5, 10, 20, 30]\,mag, $n_{\rm H}$ = [0.1, 1, 2, 10] $\times$ 10$^4$ cm$^{-3}$, T$_{kin}$ = [10, 15, 20] K, $f_{D}$ = [1, 2, 3, 5, 10], to check the equilibrium \Dntwoh \ and timescale. We adopt a constant radiation field four times stronger than the standard Habing field (G$_0$), however because of high visual extinction values considered here, small changes of G$_0$ do not affect the chemistry. We also explore initial ortho-to-para H$_2$ ratios of OPR$^{\rm H_2}_0$ = 0.001-3. A OPR$^{\rm H_2}_0$ = 3 represents the high temperature statistical ratio limit, as ortho- and para-H$_2$ are formed at high temperatures on dust grains in the ratio of their nuclear spin state statistical weights (3:1; \citealp{flower_2003}). OPR$^{\rm H_2}_0$ = 0.1-1 are values close to those deduced by \citet{crabtree_2011} in translucent clouds with T = 50-70 K: 0.8-0.3. Recently, \citet{xu_2016} made some rough constraints on the H$_2$ ortho-to-para ratio for the low density gas within Taurus. These authors estimate OPR$^{\rm H_2}\,\sim\,0.2$. Lower values are expected in dark molecular clouds (see \citealp{sipila_2013}). Therefore, our OPR$^{\rm H_2}_0$ exploration covers typical values observed in molecular clouds.

Table\,\ref{kongtable} presents the results of a subset of models with input parameters which best represent the global properties of \irdc. Shown are the equilibrium deuterium fraction, \Dntwoheq, the time taken to achieve 90\% of this value. Variation of the extinction is not displayed in the Table, as ranging A$_{v}$ between [10, 20, 30]\,mag does not significantly effect the results, however an extinction of 5\,mag tends to decrease the \Dntwoheq \ and the time by a factor of a few. This is not thought to be an issue as the mapped region of \irdc\ has an average extinction of $\sim$\,20\,mag \citep{kainulainen_2013}.

The observed deuterium fractions within \irdc \ are generally similar with the model equilibrium values, for kinetic temperatures of 10\,K and 15\,K. If we assume that the cloud has a mean global density $n_\mathrm{H}$\,$\sim$\,10$^{4}$\,cm$^{-3}$ (\citetalias{hernandez_2012a}), a mean CO depletion of $f_D$\,=\,3, and assume the mean kinetic temperature is comparable to the $\sim$\,15\,K dust temperature found by \citet{nguyen_2011}, the model predicted equilibrium value is \Dntwoheq\,=\,0.048, remarkably close to the average observed value of \Dntwoh\,=\,0.04\,$\pm$\,0.01. 

Figure\,\ref{models} shows how the ortho-to-para ratio of H$_2$ and \Dntwoh\ vary as a function of time, for the model parameters of $n_\mathrm{H}$\,=\,10$^{4}$\,cm$^{-3}$, $f_D$\,=\,3, T$_{kin}$\,= \,15\,K and A$_{v}$\,=\,20\,mag. For these properties which best describe \irdc, we find the time reach the observed deuterium fraction vary between $\sim\,0.2-8$\,Myrs, where the shortest times are for low initial ortho-to-para ratios (e.g. OPR$^{\rm H_2}_0$\,=\,0.001; also see Table\,\ref{kongtable}).

Figure\,\ref{models-OPR} displays the time needed to reach the observed deuterium fraction of \Dntwoh\,=\,0.04 from a given initial ortho-to-para ratio of H$_2$ (shown on the top axis) versus the time needed to reach this initial ortho-to-para ratio if starting from the high temperature statistical ratio limit of 3 \citep{kong_2015}. The solid blue line shows this result for the fiducial model parameters, e.g. of density, temperature, depletion factor, cosmic ray ionisation rate. The dashed blue line shows the time to reach \Dntwoh\,=\,0.02 for the fiducial model.

An estimate of the total age of the molecular cloud, i.e., from the time when the molecules formed with an assumed high temperature statistical ratio limit of 3 until the time they achieve the observed deuteration level, is the sum of the ortho-to-para ratio decay timescale and deuteration timescale: example contours of this sum for 0.3, 1, 3, 10\,Myr are shown by the dotted black lines. The astrochemical model results indicate that a timescale of at least 3\,Myrs is needed for the cloud to evolve from its initial state to the present observed deuterated state. Adopting values of CO depletion factor and density of a factor two higher ($f_D$\,=\,6 and $n_{\rm H}$\,=\,2\,$\times$\,10$^4$ cm$^{-3}$) or a cosmic ionisation rate of a factor two lower ($\zeta\,=\,1.25\,\times\,10^{-17}$\,s$^{-1}$) would imply cloud ages of at least 3-7\,Myrs, shown in lower panel of Figure\,\ref{models-OPR}. We note that decreasing the CO depletion or density, or increasing the cosmic ionisation rate would cause the models to never reach an equilibrium deuterium fraction of 0.04, hence these are not plotted on Figure\,\ref{models}. We thus conclude that the age of IRDC G035.39-00.33 is at least $\sim\,$3\,Myr. This is a lower limit, since the current observed deuteration level is consistent with astrochemical equilibrium.

 A lower limit of 3\,Myr is equivalent to $\sim$\,8 local free-fall timescales, assuming an average density of 10$^{4}$\,cm$^{-3}$ (with spherical geometry $\sim$\,4\,$\times$\,10$^{5}$\,yrs). This indicates the cloud is dynamically ``old'' and is thus likely to have had time to achieve approximate virial equilibrium (as was concluded in \citetalias{hernandez_2012a}). This timescale is consistent with that estimated from a kinematic analysis in \citetalias{henshaw_2013}. We note that future studies will involve comparison with models of evolving density (i.e. an evolving free-fall time).

Given that much of the complex structure observed in the mass surface density plot is not seen in the \ntwod \ and \ntwoh \ emission maps, it is interesting to question if we are resolving all the dense sub-structures. Core diameters extracted from 3.2\,mm continuum observations of \irdc \ with the PdBI are  typically 0.1\,pc (Henshaw et al. in preparation). This implies an approximate beam dilution factor of $\sim$\,0.07 (i.e. the square of the core-to-beam size ratio). To check this, we input typical core properties in the model. Assuming the cores have densities of $\sim$\,10$^5$\,cm$^{-3}$ (average ``core''; e.g. \citealp{butler_2012}), temperatures of $\sim$\,10\,K, and CO depletions of $\sim$\,5-10, we find model predicted \Dntwoheq \ of 0.11-0.17 (see Table\,\ref{kongtable}), which are closer to the deuteration values found in high-mass prestellar cores \citep{fontani_2011}. Applying the beam dilution factor to the predicted values, we find deuterium fractions of $\sim$\,0.007 - 0.01. Although these values are slightly below what is observed, this could be a plausible explanation for the low observed deuterium fractions in the IRDCs, where unresolved dense cores are present. The timescales to reach these deuteration levels in gas at these higher densities of 10$^{5}$\,cm$^{-3}$ are shorter: $\sim$\,1\,Myr (see also \citealp{punanova_2015}). However, this is still long compared to the local free-fall time of $\sim$\,0.1\,Myr.

Higher angular resolution observations of \ntwod\ are needed to disentangle if \irdc\ has reached a deuterium fraction equilibrium in its diffuse, $\sim$\,10$^{4}$\,cm$^{-3}$ bulk density, or if the observed deuteration is dominated by a population of denser, currently unresolved cores.




\section{Summary}\label{summary}

In this work, we have presented \ntwod\,($2-1$) data towards \irdc. The main results are summarised below: \begin{enumerate}

\item[i)] The emission from \ntwod \ is extended across \irdc, and from this emission we calculate a mean beam-averaged column density of $N$\,(\ntwod)\,=\,6.2\,$\pm$\,1.4\,$\times$\,10$^{11}$\,cm$^{-2}$. 

\item[ii)] We report an average deuterium fraction of \Dntwoh\,(\ntwod/\ntwoh)\,=\,0.04\,$\pm$\,0.01, which is three orders of magnitude higher than the interstellar [D]/[H] ratio, and within the range quoted for other IRDCs (e.g. \citealp{miettinen_2011, gerner_2015}), yet it is significantly smaller than the values found toward massive starless cores within quiescent IRDCs \citep{fontani_2011}. 

\item[iii)] We have conducted chemical modelling of the deuteration, and find that the observed values of the deuterium fraction are consistent with those of chemical equilibrium. Such an equilibrium would have taken at least $\sim$\,3\,Myr to be established. This scenario places a lower limit on the cloud age of $\sim$\,8 local free-fall times, which indicates that the IRDC filament is dynamically ``old'', with sufficient time to relax to a quasi-equilibrium virialized state. This is consistent with the previous age estimates based on the kinematics \citep{henshaw_2013}. Future studies will involve comparison with models of evolving density (i.e. an evolving free-fall time).

\item[iv)] To test if beam dilution of denser unresolved sub-structure is causing the low deuterium fraction, we input typical core properties in the model. We find that these would reach equilibrium faster (in about 1\,Myr) and have a higher equilibrium value. Using estimates for the cores size, of $\sim$\,0.1\,pc, we determine that a beam dilution factor of $\sim$\,0.07 is needed to reproduce the observed deuterium fractions, i.e. dense cold cores only occupy 7$\%$ of the cloud volume. Note that, irrespective of this, the cloud is still dynamically old.

\end{enumerate}

In light of the results, we propose that higher angular resolution observations are needed to further investigate the nature of the deuterium fraction measured across \irdc.

\section*{ACKNOWLEDGEMENTS}\label{summary}

We would like to thank the anonymous referee for their constructive comments. Furthermore, we would like to thank Michael Butler and Jouni Kainulainen for providing us with the mass surface density map, and Audra Hernandez for the CO depletion factor map. This research has made use of NASA's Astrophysics Data System. A.T.B would like to acknowledge the funding provided by Liverpool John Moores University, Max-Plank -Institute for Extraterrestrial Physics, and the University of Leeds. P.C acknowledges the financial support of the European Research Council (ERC; project PALs 320620). I.J.-S. acknowledges the funding received from the STFC through an Ernest Rutherford Fellowship (proposal number ST/L004801/1). J.C.T. acknowledges NASA grant ADAP10-0110.

\begin{table*}
\caption{Equilibrium deuterium fractions for models with an extinction of A$_{v}$\,=\,20\,mag, number densities, $n_\mathrm{H}$, of 10$^4$ and 10$^5$ cm$^{-3}$, and initial ortho-to-para H$_2$ ratios of 0.001, 0.01, 0.1, and 1. Columns show the model inputs of gas kinetic temperature, T$_{kin}$, CO depletion, and model outputs of equilibrium value of \Dntwoh and the time taken to achieve 90\,$\%$ of this value, $t_\mathrm{eq, 90}$. }
\centering
{\large
\begin{tabular}{c c c c c c c c}
\hline

T$_\mathrm{kin}$ (K) & $f_D$ & \Dntwoheq & \multicolumn{4}{c}{$t_\mathrm{eq, 90}$ (Myr)} \\ [0.5ex]
\cline{1-3}

 & & &OPR$^{\rm H_2}_0$ = 0.001 & OPR$^{\rm H_2}_0$ = 0.01 & OPR$^{\rm H_2}_0$ = 0.1 & OPR$^{\rm H_2}_0$ = 1 \\
\hline
\multicolumn{7}{c}{$n_\mathrm{H}$\,=\,10$^4$\,cm$^{-3}$ ($t_\mathrm{ff}$\,=\,4.4\,$\times$\,10$^{5}$\,yrs)}\\
\hline

10.0 & 1.0 & 0.021 & 0.73 & 3.24 & 6.19 & 8.36 \\
10.0 & 3.0 & 0.044 & 0.29 & 2.10 & 3.61 & 4.73 \\
10.0 & 5.0 & 0.057 & 0.24 & 1.79 & 3.01 & 3.92 \\
10.0 & 10.0 & 0.076 & 0.26 & 1.50 & 2.45 & 3.16 \\
\\
15.0 & 1.0 & 0.022 & 0.71 & 3.32 & 6.32 & 8.50 \\
15.0 & 3.0 & 0.048 & 0.27 & 2.06 & 3.52 & 4.60 \\
15.0 & 5.0 & 0.062 & 0.20 & 1.73 & 2.90 & 3.76 \\
15.0 & 10.0 & 0.083 & 0.19 & 1.43 & 2.32 & 2.99 \\
\\
20.0 & 1.0 & 0.015 & 0.39 & 2.91 & 6.48 & 8.72 \\
20.0 & 3.0 & 0.025 & 0.03 & 1.90 & 3.57 & 4.65 \\
20.0 & 5.0 & 0.029 & 0.003 & 1.61 & 2.91 & 3.76 \\
20.0 & 10.0 & 0.034 & 0.0005 & 1.32 & 2.32 & 2.96 \\

\hline
\multicolumn{7}{c}{$n_\mathrm{H}$\,=\,10$^5$\,cm$^{-3}$ ($t_\mathrm{ff}$\,=\,1.4\,$\times$\,10$^{5}$\,yrs)}\\
\hline

10.0 & 1.0 & 0.026 & 0.67 & 2.87 & 5.23 & 6.99 \\
10.0 & 3.0 & 0.076 & 0.45 & 1.48 & 2.35 & 3.02 \\
10.0 & 5.0 & 0.111 & 0.40 & 1.13 & 1.75 & 2.23 \\
10.0 & 10.0 & 0.17 & 0.40 & 0.84 & 1.28 & 1.61 \\
\\
15.0 & 1.0 & 0.027 & 0.65 & 2.96 & 5.44 & 7.27 \\
15.0 & 3.0 & 0.079 & 0.32 & 1.48 & 2.37 & 3.04 \\
15.0 & 5.0 & 0.115 & 0.29 & 1.12 & 1.74 & 2.21 \\
15.0 & 10.0 & 0.175 & 0.28 & 0.83 & 1.26 & 1.58 \\
\\
20.0 & 1.0 & 0.017 & 0.35 & 2.67 & 5.69 & 7.61 \\
20.0 & 3.0 & 0.033 & 0.06 & 1.45 & 2.49 & 3.18 \\
20.0 & 5.0 & 0.04 & 0.01 & 1.09 & 1.81 & 2.28 \\
20.0 & 10.0 & 0.048 & 0.0004 & 0.79 & 1.27 & 1.59 \\

\hline
\end{tabular}}
\label{kongtable}
\end{table*}

\bibliographystyle{mnras}
\bibliography{references}

\begin{thebibliography}{}
\makeatletter
\relax
\def\mn@urlcharsother{\let\do\@makeother \do\$\do\&\do\#\do\^\do\_\do\%\do\~}
\def\mn@doi{\begingroup\mn@urlcharsother \@ifnextchar [ {\mn@doi@}
  {\mn@doi@[]}}
\def\mn@doi@[#1]#2{\def\@tempa{#1}\ifx\@tempa\@empty \href
  {http://dx.doi.org/#2} {doi:#2}\else \href {http://dx.doi.org/#2} {#1}\fi
  \endgroup}
\def\mn@eprint#1#2{\mn@eprint@#1:#2::\@nil}
\def\mn@eprint@arXiv#1{\href {http://arxiv.org/abs/#1} {{\tt arXiv:#1}}}
\def\mn@eprint@dblp#1{\href {http://dblp.uni-trier.de/rec/bibtex/#1.xml}
  {dblp:#1}}
\def\mn@eprint@#1:#2:#3:#4\@nil{\def\@tempa {#1}\def\@tempb {#2}\def\@tempc
  {#3}\ifx \@tempc \@empty \let \@tempc \@tempb \let \@tempb \@tempa \fi \ifx
  \@tempb \@empty \def\@tempb {arXiv}\fi \@ifundefined
  {mn@eprint@\@tempb}{\@tempb:\@tempc}{\expandafter \expandafter \csname
  mn@eprint@\@tempb\endcsname \expandafter{\@tempc}}}

\bibitem[\protect\citeauthoryear{{Belloche}, {Parise}, {van der Tak},
  {Schilke}, {Leurini}, {G{\"u}sten}  \& {Nyman}}{{Belloche}
  et~al.}{2006}]{belloche_2006}
{Belloche} A.,  {Parise} B.,  {van der Tak} F.~F.~S.,  {Schilke} P.,  {Leurini}
  S.,  {G{\"u}sten} R.,   {Nyman} L.-{\AA}.,  2006, \mn@doi [\aap]
  {10.1051/0004-6361:20065306}, \href
  {http://adsabs.harvard.edu/abs/2006A%26A...454L..51B} {454, L51}

\bibitem[\protect\citeauthoryear{{Bergin} \& {Tafalla}}{{Bergin} \&
  {Tafalla}}{2007}]{bergin_tafalla_2007}
{Bergin} E.~A.,  {Tafalla} M.,  2007, \mn@doi [\araa]
  {10.1146/annurev.astro.45.071206.100404}, \href
  {http://adsabs.harvard.edu/abs/2007ARA%26A..45..339B} {45, 339}

\bibitem[\protect\citeauthoryear{{Bergin}, {Alves}, {Huard}  \&
  {Lada}}{{Bergin} et~al.}{2002}]{bergin_2002}
{Bergin} E.~A.,  {Alves} J.,  {Huard} T.,   {Lada} C.~J.,  2002, \mn@doi
  [\apjl] {10.1086/340950}, \href
  {http://adsabs.harvard.edu/abs/2002ApJ...570L.101B} {570, L101}

\bibitem[\protect\citeauthoryear{{Bourke}, {Myers}, {Caselli}, {Di Francesco},
  {Belloche}, {Plume}  \& {Wilner}}{{Bourke} et~al.}{2012}]{bourke_2012}
{Bourke} T.~L.,  {Myers} P.~C.,  {Caselli} P.,  {Di Francesco} J.,  {Belloche}
  A.,  {Plume} R.,   {Wilner} D.~J.,  2012, \mn@doi [\apj]
  {10.1088/0004-637X/745/2/117}, \href
  {http://adsabs.harvard.edu/abs/2012ApJ...745..117B} {745, 117}

\bibitem[\protect\citeauthoryear{{Butler} \& {Tan}}{{Butler} \&
  {Tan}}{2012}]{butler_2012}
{Butler} M.~J.,  {Tan} J.~C.,  2012, \mn@doi [\apj]
  {10.1088/0004-637X/754/1/5}, \href
  {http://adsabs.harvard.edu/abs/2012ApJ...754....5B} {754, 5}

\bibitem[\protect\citeauthoryear{{Carey} et~al.,}{{Carey}
  et~al.}{2009}]{carey_2009}
{Carey} S.~J.,  et~al., 2009, \mn@doi [\pasp] {10.1086/596581}, \href
  {http://adsabs.harvard.edu/abs/2009PASP..121...76C} {121, 76}

\bibitem[\protect\citeauthoryear{{Caselli}, {Walmsley}, {Tafalla}, {Dore}  \&
  {Myers}}{{Caselli} et~al.}{1999}]{caselli_1999}
{Caselli} P.,  {Walmsley} C.~M.,  {Tafalla} M.,  {Dore} L.,   {Myers} P.~C.,
  1999, \mn@doi [\apjl] {10.1086/312280}, \href
  {http://adsabs.harvard.edu/abs/1999ApJ...523L.165C} {523, L165}

\bibitem[\protect\citeauthoryear{{Caselli}, {Walmsley}, {Zucconi}, {Tafalla},
  {Dore}  \& {Myers}}{{Caselli} et~al.}{2002a}]{caselli_2002a}
{Caselli} P.,  {Walmsley} C.~M.,  {Zucconi} A.,  {Tafalla} M.,  {Dore} L.,
  {Myers} P.~C.,  2002a, \mn@doi [\apj] {10.1086/324302}, \href
  {http://adsabs.harvard.edu/abs/2002ApJ...565..344C} {565, 344}

\bibitem[\protect\citeauthoryear{{Caselli}, {Benson}, {Myers}  \&
  {Tafalla}}{{Caselli} et~al.}{2002b}]{caselli_2002}
{Caselli} P.,  {Benson} P.~J.,  {Myers} P.~C.,   {Tafalla} M.,  2002b, \mn@doi
  [\apj] {10.1086/340195}, \href
  {http://adsabs.harvard.edu/abs/2002ApJ...572..238C} {572, 238}

\bibitem[\protect\citeauthoryear{{Cazzoli}, {Puzzarini}  \&
  {Lapinov}}{{Cazzoli} et~al.}{2003}]{cazzoli_2003}
{Cazzoli} G.,  {Puzzarini} C.,   {Lapinov} A.~V.,  2003, \mn@doi [\apjl]
  {10.1086/377527}, \href {http://cdsads.u-strasbg.fr/abs/2003ApJ...592L..95C}
  {592, L95}

\bibitem[\protect\citeauthoryear{{Chambers}, {Jackson}, {Rathborne}  \&
  {Simon}}{{Chambers} et~al.}{2009}]{chambers_2009}
{Chambers} E.~T.,  {Jackson} J.~M.,  {Rathborne} J.~M.,   {Simon} R.,  2009,
  \mn@doi [\apjs] {10.1088/0067-0049/181/2/360}, \href
  {http://adsabs.harvard.edu/abs/2009ApJS..181..360C} {181, 360}

\bibitem[\protect\citeauthoryear{{Clemens}}{{Clemens}}{1985}]{clemens_1985}
{Clemens} D.~P.,  1985, \mn@doi [\apj] {10.1086/163386}, \href
  {http://adsabs.harvard.edu/abs/1985ApJ...295..422C} {295, 422}

\bibitem[\protect\citeauthoryear{{Crabtree}, {Indriolo}, {Kreckel}, {Tom}  \&
  {McCall}}{{Crabtree} et~al.}{2011}]{crabtree_2011}
{Crabtree} K.~N.,  {Indriolo} N.,  {Kreckel} H.,  {Tom} B.~A.,   {McCall}
  B.~J.,  2011, \mn@doi [\apj] {10.1088/0004-637X/729/1/15}, \href
  {http://adsabs.harvard.edu/abs/2011ApJ...729...15C} {729, 15}

\bibitem[\protect\citeauthoryear{{Crapsi}, {Caselli}, {Walmsley}, {Tafalla},
  {Lee}, {Bourke}  \& {Myers}}{{Crapsi} et~al.}{2004}]{crapsi_2004}
{Crapsi} A.,  {Caselli} P.,  {Walmsley} C.~M.,  {Tafalla} M.,  {Lee} C.~W.,
  {Bourke} T.~L.,   {Myers} P.~C.,  2004, \mn@doi [\aap]
  {10.1051/0004-6361:20035915}, \href
  {http://adsabs.harvard.edu/abs/2004A%26A...420..957C} {420, 957}

\bibitem[\protect\citeauthoryear{{Crapsi} et~al.,}{{Crapsi}
  et~al.}{2005}]{crapsi_2005}
{Crapsi} A.,  et~al., 2005, \mn@doi [\aap] {10.1051/0004-6361:20042411}, \href
  {http://adsabs.harvard.edu/abs/2005A%26A...439.1023C} {439, 1023}

\bibitem[\protect\citeauthoryear{{Dalgarno} \& {Lepp}}{{Dalgarno} \&
  {Lepp}}{1984}]{dalgarno_1984}
{Dalgarno} A.,  {Lepp} S.,  1984, \mn@doi [\apjl] {10.1086/184395}, \href
  {http://adsabs.harvard.edu/abs/1984ApJ...287L..47D} {287, L47}

\bibitem[\protect\citeauthoryear{{Dore}, {Caselli}, {Beninati}, {Bourke},
  {Myers}  \& {Cazzoli}}{{Dore} et~al.}{2004}]{dore_2004}
{Dore} L.,  {Caselli} P.,  {Beninati} S.,  {Bourke} T.,  {Myers} P.~C.,
  {Cazzoli} G.,  2004, \mn@doi [\aap] {10.1051/0004-6361:20034025}, \href
  {http://adsabs.harvard.edu/abs/2004A%26A...413.1177D} {413, 1177}

\bibitem[\protect\citeauthoryear{{Flower}}{{Flower}}{2003}]{flower_2003}
{Flower} D.,  2003, {Molecular Collisions in the Interstellar Medium}

\bibitem[\protect\citeauthoryear{{Fontani}, {Caselli}, {Crapsi}, {Cesaroni},
  {Molinari}, {Testi}  \& {Brand}}{{Fontani} et~al.}{2006}]{fontani_2006}
{Fontani} F.,  {Caselli} P.,  {Crapsi} A.,  {Cesaroni} R.,  {Molinari} S.,
  {Testi} L.,   {Brand} J.,  2006, \mn@doi [\aap] {10.1051/0004-6361:20066105},
  \href {http://adsabs.harvard.edu/abs/2006A%26A...460..709F} {460, 709}

\bibitem[\protect\citeauthoryear{{Fontani} et~al.,}{{Fontani}
  et~al.}{2011}]{fontani_2011}
{Fontani} F.,  et~al., 2011, \mn@doi [\aap] {10.1051/0004-6361/201116631},
  \href {http://adsabs.harvard.edu/abs/2011A%26A...529L...7F} {529, L7+}

\bibitem[\protect\citeauthoryear{{Fontani}, {Palau}, {Busquet}, {Isella},
  {Estalella}, {Sanchez-Monge}, {Caselli}  \& {Zhang}}{{Fontani}
  et~al.}{2012a}]{fontani_2012b}
{Fontani} F.,  {Palau} A.,  {Busquet} G.,  {Isella} A.,  {Estalella} R.,
  {Sanchez-Monge} {\'A}.,  {Caselli} P.,   {Zhang} Q.,  2012a, \mn@doi [\mnras]
  {10.1111/j.1365-2966.2012.20990.x}, \href
  {http://adsabs.harvard.edu/abs/2012MNRAS.423.1691F} {423, 1691}

\bibitem[\protect\citeauthoryear{{Fontani}, {Giannetti}, {Beltr{\'a}n},
  {Dodson}, {Rioja}, {Brand}, {Caselli}  \& {Cesaroni}}{{Fontani}
  et~al.}{2012b}]{fontani_2012a}
{Fontani} F.,  {Giannetti} A.,  {Beltr{\'a}n} M.~T.,  {Dodson} R.,  {Rioja} M.,
   {Brand} J.,  {Caselli} P.,   {Cesaroni} R.,  2012b, \mn@doi [\mnras]
  {10.1111/j.1365-2966.2012.21043.x}, \href
  {http://adsabs.harvard.edu/abs/2012MNRAS.423.2342F} {423, 2342}

\bibitem[\protect\citeauthoryear{{Fontani}, {Caselli}, {Zhang}, {Brand},
  {Busquet}  \& {Palau}}{{Fontani} et~al.}{2012c}]{fontani_2012c}
{Fontani} F.,  {Caselli} P.,  {Zhang} Q.,  {Brand} J.,  {Busquet} G.,   {Palau}
  A.,  2012c, \mn@doi [\aap] {10.1051/0004-6361/201118153}, \href
  {http://adsabs.harvard.edu/abs/2012A%26A...541A..32F} {541, A32}

\bibitem[\protect\citeauthoryear{{Friesen}, {Di Francesco}, {Myers},
  {Belloche}, {Shirley}, {Bourke}  \& {Andr{\'e}}}{{Friesen}
  et~al.}{2010}]{friesen_2010}
{Friesen} R.~K.,  {Di Francesco} J.,  {Myers} P.~C.,  {Belloche} A.,  {Shirley}
  Y.~L.,  {Bourke} T.~L.,   {Andr{\'e}} P.,  2010, \mn@doi [\apj]
  {10.1088/0004-637X/718/2/666}, \href
  {http://adsabs.harvard.edu/abs/2010ApJ...718..666F} {718, 666}

\bibitem[\protect\citeauthoryear{{Friesen}, {Kirk}  \& {Shirley}}{{Friesen}
  et~al.}{2013}]{friesen_2013}
{Friesen} R.~K.,  {Kirk} H.~M.,   {Shirley} Y.~L.,  2013, \mn@doi [\apj]
  {10.1088/0004-637X/765/1/59}, \href
  {http://adsabs.harvard.edu/abs/2013ApJ...765...59F} {765, 59}

\bibitem[\protect\citeauthoryear{{Gerner}, {Shirley}, {Beuther}, {Semenov},
  {Linz}, {Abertsson}  \& {Henning}}{{Gerner} et~al.}{2015}]{gerner_2015}
{Gerner} T.,  {Shirley} Y.,  {Beuther} H.,  {Semenov} D.,  {Linz} H.,
  {Abertsson} T.,   {Henning} T.,  2015, preprint, \href
  {http://adsabs.harvard.edu/abs/2015arXiv150306594G} {} (\mn@eprint {arXiv}
  {1503.06594})

\bibitem[\protect\citeauthoryear{{Giannetti} et~al.,}{{Giannetti}
  et~al.}{2014}]{giannetti_2014}
{Giannetti} A.,  et~al., 2014, \mn@doi [\aap] {10.1051/0004-6361/201423692},
  \href {http://adsabs.harvard.edu/abs/2014A%26A...570A..65G} {570, A65}

\bibitem[\protect\citeauthoryear{{Henshaw}, {Caselli}, {Fontani},
  {Jim{\'e}nez-Serra}, {Tan}  \& {Hernandez}}{{Henshaw}
  et~al.}{2013}]{henshaw_2013}
{Henshaw} J.~D.,  {Caselli} P.,  {Fontani} F.,  {Jim{\'e}nez-Serra} I.,  {Tan}
  J.~C.,   {Hernandez} A.~K.,  2013, \mn@doi [\mnras] {10.1093/mnras/sts282},
  \href {http://adsabs.harvard.edu/abs/2013MNRAS.428.3425H} {428, 3425}

\bibitem[\protect\citeauthoryear{{Henshaw}, {Caselli}, {Fontani},
  {Jim{\'e}nez-Serra}  \& {Tan}}{{Henshaw} et~al.}{2014}]{henshaw_2014}
{Henshaw} J.~D.,  {Caselli} P.,  {Fontani} F.,  {Jim{\'e}nez-Serra} I.,   {Tan}
  J.~C.,  2014, \mn@doi [\mnras] {10.1093/mnras/stu446}, \href
  {http://adsabs.harvard.edu/abs/2014MNRAS.440.2860H} {440, 2860}

\bibitem[\protect\citeauthoryear{{Hernandez}, {Tan}, {Caselli}, {Butler},
  {Jim{\'e}nez-Serra}, {Fontani}  \& {Barnes}}{{Hernandez}
  et~al.}{2011}]{hernandez_2011}
{Hernandez} A.~K.,  {Tan} J.~C.,  {Caselli} P.,  {Butler} M.~J.,
  {Jim{\'e}nez-Serra} I.,  {Fontani} F.,   {Barnes} P.,  2011, \mn@doi [\apj]
  {10.1088/0004-637X/738/1/11}, \href
  {http://cdsads.u-strasbg.fr/abs/2011ApJ...738...11H} {738, 11}

\bibitem[\protect\citeauthoryear{{Hernandez}, {Tan}, {Kainulainen}, {Caselli},
  {Butler}, {Jim{\'e}nez-Serra}  \& {Fontani}}{{Hernandez}
  et~al.}{2012}]{hernandez_2012a}
{Hernandez} A.~K.,  {Tan} J.~C.,  {Kainulainen} J.,  {Caselli} P.,  {Butler}
  M.~J.,  {Jim{\'e}nez-Serra} I.,   {Fontani} F.,  2012, \mn@doi [\apjl]
  {10.1088/2041-8205/756/1/L13}, \href
  {http://adsabs.harvard.edu/abs/2012ApJ...756L..13H} {756, L13}

\bibitem[\protect\citeauthoryear{{Jim{\'e}nez-Serra}, {Caselli}, {Tan},
  {Hernandez}, {Fontani}, {Butler}  \& {van Loo}}{{Jim{\'e}nez-Serra}
  et~al.}{2010}]{izaskun_2010}
{Jim{\'e}nez-Serra} I.,  {Caselli} P.,  {Tan} J.~C.,  {Hernandez} A.~K.,
  {Fontani} F.,  {Butler} M.~J.,   {van Loo} S.,  2010, \mn@doi [\mnras]
  {10.1111/j.1365-2966.2010.16698.x}, \href
  {http://adsabs.harvard.edu/abs/2010MNRAS.406..187J} {406, 187}

\bibitem[\protect\citeauthoryear{{Jim{\'e}nez-Serra}, {Caselli}, {Fontani},
  {Tan}, {Henshaw}, {Kainulainen}  \& {Hernandez}}{{Jim{\'e}nez-Serra}
  et~al.}{2014}]{jimenez_2014}
{Jim{\'e}nez-Serra} I.,  {Caselli} P.,  {Fontani} F.,  {Tan} J.~C.,  {Henshaw}
  J.~D.,  {Kainulainen} J.,   {Hernandez} A.~K.,  2014, \mn@doi [\mnras]
  {10.1093/mnras/stu078}, \href
  {http://adsabs.harvard.edu/abs/2014MNRAS.439.1996J} {439, 1996}

\bibitem[\protect\citeauthoryear{{Kainulainen} \& {Tan}}{{Kainulainen} \&
  {Tan}}{2013}]{kainulainen_2013}
{Kainulainen} J.,  {Tan} J.~C.,  2013, \mn@doi [\aap]
  {10.1051/0004-6361/201219526}, \href
  {http://adsabs.harvard.edu/abs/2013A%26A...549A..53K} {549, A53}

\bibitem[\protect\citeauthoryear{{Kong}, {Caselli}, {Tan}, {Wakelam}  \&
  {Sipil{\"a}}}{{Kong} et~al.}{2015}]{kong_2015}
{Kong} S.,  {Caselli} P.,  {Tan} J.~C.,  {Wakelam} V.,   {Sipil{\"a}} O.,
  2015, \mn@doi [\apj] {10.1088/0004-637X/804/2/98}, \href
  {http://adsabs.harvard.edu/abs/2015ApJ...804...98K} {804, 98}

\bibitem[\protect\citeauthoryear{{Kong} et~al.,}{{Kong}
  et~al.}{2016}]{kong_2016}
{Kong} S.,  et~al., 2016, \apj, \href
  {http://adsabs.harvard.edu/abs/2015arXiv150908684K} {}

\bibitem[\protect\citeauthoryear{{Kutner} \& {Ulich}}{{Kutner} \&
  {Ulich}}{1981}]{kutner_1981}
{Kutner} M.~L.,  {Ulich} B.~L.,  1981, \mn@doi [\apj] {10.1086/159380}, \href
  {http://adsabs.harvard.edu/abs/1981ApJ...250..341K} {250, 341}

\bibitem[\protect\citeauthoryear{{Le Petit}, {Roueff}  \& {Le Bourlot}}{{Le
  Petit} et~al.}{2002}]{lepetit_2002}
{Le Petit} F.,  {Roueff} E.,   {Le Bourlot} J.,  2002, \mn@doi [\aap]
  {10.1051/0004-6361:20020729}, \href
  {http://adsabs.harvard.edu/abs/2002A%26A...390..369L} {390, 369}

\bibitem[\protect\citeauthoryear{{Miettinen}, {Hennemann}  \&
  {Linz}}{{Miettinen} et~al.}{2011}]{miettinen_2011}
{Miettinen} O.,  {Hennemann} M.,   {Linz} H.,  2011, \mn@doi [\aap]
  {10.1051/0004-6361/201117187}, \href
  {http://adsabs.harvard.edu/abs/2011A%26A...534A.134M} {534, A134}

\bibitem[\protect\citeauthoryear{{Millar}, {Bennett}  \& {Herbst}}{{Millar}
  et~al.}{1989}]{millar_1989}
{Millar} T.~J.,  {Bennett} A.,   {Herbst} E.,  1989, \mn@doi [\apj]
  {10.1086/167444}, \href {http://adsabs.harvard.edu/abs/1989ApJ...340..906M}
  {340, 906}

\bibitem[\protect\citeauthoryear{{Nguyen Luong} et~al.,}{{Nguyen Luong}
  et~al.}{2011}]{nguyen_2011}
{Nguyen Luong} Q.,  et~al., 2011, \mn@doi [\aap] {10.1051/0004-6361/201117831},
  \href {http://adsabs.harvard.edu/abs/2011A%26A...535A..76N} {535, A76}

\bibitem[\protect\citeauthoryear{{Oliveira}, {H{\'e}brard}, {Howk}, {Kruk},
  {Chayer}  \& {Moos}}{{Oliveira} et~al.}{2003}]{oliveira_2003}
{Oliveira} C.~M.,  {H{\'e}brard} G.,  {Howk} J.~C.,  {Kruk} J.~W.,  {Chayer}
  P.,   {Moos} H.~W.,  2003, \mn@doi [\apj] {10.1086/368019}, \href
  {http://adsabs.harvard.edu/abs/2003ApJ...587..235O} {587, 235}

\bibitem[\protect\citeauthoryear{{Pagani}, {Salez}  \& {Wannier}}{{Pagani}
  et~al.}{1992}]{pagani_1992}
{Pagani} L.,  {Salez} M.,   {Wannier} P.~G.,  1992, \aap, \href
  {http://adsabs.harvard.edu/abs/1992A%26A...258..479P} {258, 479}

\bibitem[\protect\citeauthoryear{{Pagani}, {Bacmann}, {Cabrit}  \&
  {Vastel}}{{Pagani} et~al.}{2007}]{pagani_2007}
{Pagani} L.,  {Bacmann} A.,  {Cabrit} S.,   {Vastel} C.,  2007, \mn@doi [\aap]
  {10.1051/0004-6361:20066670}, \href
  {http://cdsads.u-strasbg.fr/abs/2007A%26A...467..179P} {467, 179}

\bibitem[\protect\citeauthoryear{{Pagani} et~al.,}{{Pagani}
  et~al.}{2009a}]{pagani_2009b}
{Pagani} L.,  et~al., 2009a, \mn@doi [\aap] {10.1051/0004-6361:200810587},
  \href {http://adsabs.harvard.edu/abs/2009A%26A...494..623P} {494, 623}

\bibitem[\protect\citeauthoryear{{Pagani}, {Daniel}  \& {Dubernet}}{{Pagani}
  et~al.}{2009b}]{pagani_2009}
{Pagani} L.,  {Daniel} F.,   {Dubernet} M.,  2009b, \mn@doi [\aap]
  {10.1051/0004-6361:200810570}, \href
  {http://adsabs.harvard.edu/abs/2009A%26A...494..719P} {494, 719}

\bibitem[\protect\citeauthoryear{{Punanova}, {Caselli}, {Pon}, {Belloche}  \&
  {Andr{\'e}}}{{Punanova} et~al.}{2015}]{punanova_2015}
{Punanova} A.,  {Caselli} P.,  {Pon} A.,  {Belloche} A.,   {Andr{\'e}} P.,
  2015, preprint, \href {http://adsabs.harvard.edu/abs/2015arXiv151202986P} {}
  (\mn@eprint {arXiv} {1512.02986})

\bibitem[\protect\citeauthoryear{{Ragan}, {Bergin}  \& {Wilner}}{{Ragan}
  et~al.}{2011}]{ragan_2011}
{Ragan} S.~E.,  {Bergin} E.~A.,   {Wilner} D.,  2011, \mn@doi [\apj]
  {10.1088/0004-637X/736/2/163}, \href
  {http://cdsads.u-strasbg.fr/abs/2011ApJ...736..163R} {736, 163}

\bibitem[\protect\citeauthoryear{{Rathborne}, {Jackson}  \&
  {Simon}}{{Rathborne} et~al.}{2006}]{rathborne_2006}
{Rathborne} J.~M.,  {Jackson} J.~M.,   {Simon} R.,  2006, \mn@doi [\apj]
  {10.1086/500423}, \href {http://adsabs.harvard.edu/abs/2006ApJ...641..389R}
  {641, 389}

\bibitem[\protect\citeauthoryear{{Simon}, {Rathborne}, {Shah}, {Jackson}  \&
  {Chambers}}{{Simon} et~al.}{2006}]{simon_2006b}
{Simon} R.,  {Rathborne} J.~M.,  {Shah} R.~Y.,  {Jackson} J.~M.,   {Chambers}
  E.~T.,  2006, \mn@doi [\apj] {10.1086/508915}, \href
  {http://adsabs.harvard.edu/abs/2006ApJ...653.1325S} {653, 1325}

\bibitem[\protect\citeauthoryear{{Sipil{\"a}}, {Caselli}  \&
  {Harju}}{{Sipil{\"a}} et~al.}{2013}]{sipila_2013}
{Sipil{\"a}} O.,  {Caselli} P.,   {Harju} J.,  2013, \mn@doi [\aap]
  {10.1051/0004-6361/201220922}, \href
  {http://adsabs.harvard.edu/abs/2013A%26A...554A..92S} {554, A92}

\bibitem[\protect\citeauthoryear{{Tan}, {Beltr{\'a}n}, {Caselli}, {Fontani},
  {Fuente}, {Krumholz}, {McKee}  \& {Stolte}}{{Tan} et~al.}{2014}]{tan_2014}
{Tan} J.~C.,  {Beltr{\'a}n} M.~T.,  {Caselli} P.,  {Fontani} F.,  {Fuente} A.,
  {Krumholz} M.~R.,  {McKee} C.~F.,   {Stolte} A.,  2014, \mn@doi [Protostars
  and Planets VI] {10.2458/azu_uapress_9780816531240-ch007}, \href
  {http://adsabs.harvard.edu/abs/2014prpl.conf..149T} {pp 149--172}

\bibitem[\protect\citeauthoryear{{Wakelam} et~al.,}{{Wakelam}
  et~al.}{2012}]{wakelam_2012}
{Wakelam} V.,  et~al., 2012, \mn@doi [\apjs] {10.1088/0067-0049/199/1/21},
  \href {http://adsabs.harvard.edu/abs/2012ApJS..199...21W} {199, 21}

\bibitem[\protect\citeauthoryear{{Walmsley}, {Flower}  \& {Pineau des
  For{\^e}ts}}{{Walmsley} et~al.}{2004}]{walmsley_2004}
{Walmsley} C.~M.,  {Flower} D.~R.,   {Pineau des For{\^e}ts} G.,  2004, \mn@doi
  [\aap] {10.1051/0004-6361:20035718}, \href
  {http://adsabs.harvard.edu/abs/2004A%26A...418.1035W} {418, 1035}

\bibitem[\protect\citeauthoryear{{Xu}, {Li}, {Yue}  \& {Goldsmith}}{{Xu}
  et~al.}{2016}]{xu_2016}
{Xu} D.,  {Li} D.,  {Yue} N.,   {Goldsmith} P.~F.,  2016, preprint, \href
  {http://adsabs.harvard.edu/abs/2016arXiv160103165X} {} (\mn@eprint {arXiv}
  {1601.03165})

\bibitem[\protect\citeauthoryear{{di Francesco}, {Evans}, {Caselli}, {Myers},
  {Shirley}, {Aikawa}  \& {Tafalla}}{{di Francesco}
  et~al.}{2007}]{difrancrsco_2007}
{di Francesco} J.,  {Evans} II N.~J.,  {Caselli} P.,  {Myers} P.~C.,  {Shirley}
  Y.,  {Aikawa} Y.,   {Tafalla} M.,  2007, Protostars and Planets V, \href
  {http://adsabs.harvard.edu/abs/2007prpl.conf...17D} {pp 17--32}

\makeatother
\end{thebibliography}





\bsp	
\label{lastpage}
\end{document}